\documentstyle[12pt]{article}
\newcommand{\be}{\begin{equation}}
\newcommand{\ee}{\end{equation}}
\newcommand{\lpa}{\lambda_{\parallel}}
\newcommand{\lpe}{\lambda_{\perp}}
\input{jour.tex}
\begin{document}
\baselineskip 0.8cm
\hfill {\tt SNUTP 93--28}
\begin{center}
{\LARGE\bf Dynamics of Toom interface \\ in three dimensions \\}
\vspace{1.0cm}
{H.~Jeong$^{\dag}$, B.~Kahng$^{\ddag}$, and D.~Kim$^{\dag}$ \\}
\vspace{1.0cm}
{\sl \dag Center for Theoretical Physics and Department of Physics, \\
Seoul National University, Seoul 151-742, Korea \\}
\vspace{0.5cm}
{\sl \ddag Department of Physics, Kon-Kuk University, Seoul 133-701, Korea \\}
\vspace{1.0cm}
{\LARGE Abstract}
\end{center}
We introduce a novel three dimensional Toom model on bcc lattice, and
study its physical properties. In low-noise limit, the model leads
to an effective solid-on-solid-type model, which exhibits
a stationary interface via depositions and evaporations
with avalanche process. We find that the model is described by
the Edwards-Wilkinson equation for unbiased case and the
anisotropic-Kardar-Parisi-Zhang equation in the $weak$-coupling
limit for biased case. Thus the square of the surface width
diverges logarithmically with space and time for both unbiased and
biased cases.\\


\newpage
In the past few years, there has been an explosion of studies in the
field of non-equilibrium surface growth due to
theoretical interests in the classification of universality for
stochastic models and also due to the application to physical
phenomena such as crystal growth, vapor deposition, electroplating,
biological growth etc [1]. Recently Derrida, Lebowitz, Speer, and Spohn
(DLSS) [2] studied physical properties of the two dimensional Toom model.
The Toom interface is formed by a rule of simple probabilistic
cellular automaton.
In low-noise limit, this model leads
to a ($1+1$) dimensional solid-on-solid-type (SOS) model,
which is in turn much simpler for understanding
generic nature of dynamics.
In the SOS model, the dynamics of spin-flips may be regarded as
a deposition-evaporation process of particles. Due to the
nature of the Toom dynamics, the deposition-evaporation process occurs in
an avalanche fashion.

Extension of the Toom model into three dimensions
is an interesting problem. However it is not obvious
to define a majority rule on $simple$ $cubic$ (sc) lattice.
The majority rule applying to the group of spins
consisting of itself and the three nearest-neighbor spins in
each direction as in the original square lattice is ambiguous,
because even number of spins are involved.
Recently Barab\'asi, Araujo, and Stanley (BAS) [3] overcame this
difficulty, and introduced a majority rule on sc structure
using the five spins
consisting of the next-nearest-neighbor spin in the
$(-1,-1,1)$-direction in addition to the four spins mentioned above.
Including the fifth spin in that particular direction leads
to the anisotropic Kardar-Parisi-Zhang (AKPZ) equation [4]
for the interface,
\be
\partial_t h=\nu_{\parallel}{{\partial}_{\parallel}}^2 h +
\nu_{\perp}{{\partial}_{\perp}}^2 h
+{1\over 2}\lpa(\partial_{\parallel}h)^2
+{1\over 2}\lpe(\partial_{\perp}h)^2+\eta
\ee
with white noise $\eta$.
BAS found that their model belongs to the $strong$-coupling regime
of the AKPZ universality, which is equivalent to the isotropic KPZ
universality [5]. Thus the square of the surface width shows a
power-law type divergence.\\

In this letter, we introduce a new three dimensional Toom model on
$body$ $centered$ $cubic$ ($bcc$) lattice, in which majority rule
is applied to the set of five spins comprising of
itself and four $nearest$ neighboring spins as depicted in Fig.~1.
Our model is distinguished from that of BAS in that the
four neighboring spins
are of equidistance, while they are not in the BAS model.
Interestingly, this difference of the local dynamic rules leads
to different universality classes, which is in accordance with a
recent claim that dynamic universality class
depends on local dynamic rules in (2+1)-dimensions [6].
We studied the model for both unbiased and biased cases, and
found that our model belongs to the Edwards-Wilkinson (EW) [7] universality
for the unbiased case and the AKPZ universality
in the $weak$-coupling limit [4]
for the biased case. The latter is shown by demonstrating that the signs
of the coefficients $\lambda_{\parallel}$ and $\lambda_{\perp}$ are
opposite to each other.
Therefore our model is in a sense a completely opposite limit
in comparison to the BAS model.
Consequently, the square of the surface width exhibits
a logarithmic divergence with space and time
for both biased and unbiased cases. As far as we are aware of,
this is the first stochastic lattice model which exhibits opposite signs for
$\lambda_{\parallel}$ and $\lambda_{\perp}$\\

The model we introduce here
consists of Ising spins ($\sigma_{i,j,k}=\pm 1$) on bcc lattice.
At each time, a randomly selected spin is updated according to
the local rule that $\sigma_{i,j,k}(t+1)$ becomes, with probability $1-p-q$,
equal to the majority of itself and
the four nearest-neighbor spins at
$(i\pm \frac{1}{2},j-\frac{1}{2},k\pm \frac{1}{2})$ at time $t$,
and otherwise becomes equal to $+1$ with probability $p$, and $-1$ with
probability $q$.
Unbiased (biased) dynamics results when $p=q$ ($p\ne q$).
Besides this dynamic rule, we use proper boundary conditions
to generate reasonable interface between $(+)$ and $(-)$ spin domains.
The boundary spins on the plane $j=0$ is fixed to the value,
$\sigma_{i,0,k}= +1~ (-1)$ if $k > L/2$~ ($k < L/2$).
Periodic boundary conditions are imposed on other boundary planes.
With these boundary conditions, the Toom interface in three dimensions
is formed between upper $(+)$ spin domain and lower $(-)$ spin domain.\\

In the low-noise limit ($p, q \rightarrow 0$), a spin flipped due
to noise returns immediately to its original state by the majority rule, if
the spin is situated away from the interface.
Accordingly, we may assume that spin flips occur only at the boundary.
We are thus led to study an effective SOS-type
model in analogy with the (1+1)-dimensional stairlike model studied
by DLSS. In the SOS model,
the dynamic rule of spin-flips is mapped to a particle dynamics on
a two dimensional substrate in the form of deposition-evaporation
with avalanche. If the avalanche process is not allowed, so that
depositions and evaporations occur only on local valleys and mountains
respectively, then our model would be equivalent to
the deposition-evaporation model proposed by Forrest and Tang [8],
a generalization of the Plischke-R\'acz-Liu model [9]
into higher dimensions.\\

The SOS model is defined on the checkerboard lattice, square lattice
rotated by 45 degree.
To each lattice point, a relative height value is assigned.
The value represents the height of the interface of the original
three dimensional Toom model. Initially we begin with a flat surface
characterized by the heights 0 on one sublattice and 1 on the
other (see Fig.~2).
The Toom dynamics is then translated to following updating rule:
At each time step, we select a random site, ($i,j$) and start evaporation
(deposition) process with probability $\bar p$ (probability $1-\bar p$).
In the evaporation (deposition) process, the height of the site is
decreased (increased)  by 2 if the heights of both of its two
nearest neighbors at $(i+\frac{1}{2},j-\frac{1}{2})$ and
$(i-\frac{1}{2},j-\frac{1}{2})$ are lower (higher).
Otherwise there is no change.
Next the avalanche process may occur on the sites,
$(i+\frac{1}{2},j+\frac{1}{2})$ and $(i-\frac{1}{2},j+\frac{1}{2})$.
If the height of each site is higher (lower) by 3 than
that of $(i,j)$, then the height is decreased (increased) by 2.
The avalanche rule is then applied successively to next rows in
$\hat{j}$-direction until there is no change.
The avalanche process is an interesting aspect of our model,
and reflects a generic feature of the Toom dynamics.
The unbiased case $p=q$ in bcc structure
corresponds to the case of $\bar p=0.5$ in the SOS model,
where deposition and evaporation occur with equal probability.
The biased case $p\ne q$ corresponds to the case of $\bar p\ne 0.5$
in the SOS model.
We impose periodic boundary conditions in both directions.
\\

The continuum equation for our model is also described by Eq.~(1),
because our model selects out a preferred direction, and because the
cubic nonlinear term derived by DLSS was proved to be marginally
irrelevant even in (1+1) dimensions [10].
For the unbiased case, both of the nonlinear terms in Eq.~(1)
disappear due to the symmetry of deposition and evaporation.
So the equation reduces to the EW equation implying that
the square of the surface width diverges logarithmically with space
and time. For the biased case, average height
grows with increasing time, so that $\lpa$ and $\lpe$ are
nonzero. In order to find out the signs of $\lpa$ and $\lpe$, we consider
the following. \\

For convenience, let us consider the pure deposition case ($\bar p=0$)
on three kinds of substrates with linear size $L$;
(a) a flat substrate, (b) a substrate with one step along a row,
and (c) a substrate with one step along a column. The case of (b)
is shown in Fig.~2. For the case of (a), the average number of
particles deposited in one time step is $L^2\over 2L^2$, since only
local valleys are available for deposition. Thus the average height
change is $\big<\Delta h \big>_0={L^2\over 2L^2}\cdot 2 =1$.
For the case of (b), the average height changes
by $\big<\Delta h \big>_{\parallel}={L(L-1)\over {2L^2}}\cdot 2
+ {L\over {2L^2}} \cdot 6=1+{2\over L}$, where
the first term results from deposition at local valleys and the
second one comes from the avalanche on the downward hill.
Therefore, $\big<\Delta h \big>_{\parallel} > \big<\Delta h
\big>_0$, which implies $\lpa > 0 $.
Next, for the case of $(c)$, the average height change
is $\big< \Delta h \big>_{\perp}={L(L-1) \over {2L^2}}\cdot
2=1-{1\over L}$, because there is one less column of valleys for deposition.
Therefore, $\big<\Delta h \big>_{\perp } < \big<\Delta h \big>_0$,
which implies $\lpe < 0 $. Above considerations can be generalized
to initial substrates with arbitrary slope and to arbitrary $\bar p \ne 0$.
Since the signs of $\lambda_{\parallel}$ and $\lambda_{\perp}$
are opposite, the AKPZ equation renormalizes to
the weak-coupling limit as shown by Wolf [4].
Consequently, our model belongs to the weak-coupling regime of
the AKPZ universality. Therefore, the
square of the surface width is logarithmic for both unbiased and
biased cases.\\

We performed numerical simulations of the three dimensional
Toom model on bcc lattice, and compared the result with that of
the ($2+1$)-dimensional SOS model. Results for both cases run
on small sizes are in complete agreement with each other
in low-noise limit. Accordingly, we performed
simulations intensively for larger systems using the SOS model.
The simulations are done in the range of system size
$L=20\sim 140$ for several values of $\bar p$.
For all cases, we expect that the square of the surface
width $w^2$ diverges logarithmically with space and time as $w^2 \sim \ln t$
before saturation, and $w^2 \sim \ln L$ after saturation.
We present the result for $\bar p=0.5$ (unbiased) and $\bar p=0.3$
(biased) in Fig.~3. The numerical data are in good agreement
with the theoretical predictions. Besides the square of the
surface width, we also measured the height-height correlation functions
in $\hat {i}-$ and $\hat{j}-$directions, respectively, and
found that they behave in the same manner as $w^2$.
The detail numerical data will be presented elsewhere. [11]\\

Finally we examined the avalanche size distribution
$D(s)$, which is defined as the number of successive spin flips
by a single noise process. $D(s)$ was measured
in two different manners. In the first case, it is measured in the
critical state (after saturation), while in the second case, it is
measured during the whole time steps. In both cases,
the distribution function $D(s)$ is found to be exponential,
$D(s)\sim \exp(-s/s^*)$ with a constant value $s^*=1.097\pm 0.057$ [11].
The exponential-type functional form seems to reflect indirectly
the validity of the collective-variable approximation used by DLSS [2].\\

In conclusion, we have introduced a new three dimensional Toom model,
and its associated SOS-type model in $(2+1)$-dimension. The spin
dynamics in three dimension is mapped into particle dynamics via
deposition and evaporation process with avalanche
on the checkerboard lattice.
We have found that for the unbiased case, the interface is described
by the EW equation, and for the biased case, it is described
by the AKPZ equation with the opposite signs
of $\lambda_{\parallel}$ and $\lambda_{\perp}$.
Consequently, the square of the surface width diverges logarithmically
with space and time. This result is confirmed by numerical simulations.\\

We would like to thank Dr.~Jin~Min~Kim for helpful discussions.
This work was supported in part by the KOSEF through the SRC
program of SNU-CTP and through CTSP in Korea Univ, and in part
by the Ministry of Education, Korea. \\

\newpage
{\Large\bf Figure Captions }
\begin{description}
\item[Fig.~1 ] The three dimensional Toom rule on bcc lattice used
in this work. The black circled spin $\sigma_{i,j,k}$ is updated with
the majority rule of itself and four nearest neighbor spins
(the black circled).
\item[Fig.~2 ] The configuration of a substrate with one step along a
row for linear size $L=4$.
\item[Fig.~3 ] $w^2$ versus $\ln t$ for unbiased case $\bar p =0.5$ (a)
and for biased case $\bar p=0.3$ (b).
Each curve is for system sizes $L=20, 40, 60, 80, 100, 120$ and $140$,
respectively,  from bottom to top, and is averaged over 300 configurations.
Insets: $w^2$ versus $\ln L$ after saturation for each case.
\end{description}
\end{document}